\documentclass[aps,prl,superscriptaddress,twocolumn]{revtex4-1}

\usepackage[pdftex]{graphicx,xcolor}
\usepackage{amsmath}
\usepackage{amsfonts}
\usepackage{amssymb}
\usepackage{mathrsfs}
\usepackage{bbm}
\usepackage{bbold}
\usepackage{subfigure}

\newcommand {\be}{\begin{eqnarray}}
\newcommand {\ee}{\end{eqnarray}}

\newcommand{\bk}{{\bf k}}

\newcommand{\bq}{{\bf q}}
\newcommand{\br}{{\bf r}}

\newcommand{\tst}{\textstyle}

\newcommand{\half}{\frac{1}{2}}

\begin{document}

\title {Theory of Valley-Density Wave and Hidden Order in Iron-Pnictides}
\author {Jian Kang}
\affiliation {Institute for Quantum Matter and Department
of Physics \& Astronomy, The Johns Hopkins University, Baltimore, MD 21218}
\author {Zlatko Te\v sanovi{\' c}}
\affiliation {Institute for Quantum Matter and Department
of Physics \& Astronomy, The Johns Hopkins University, Baltimore, MD 21218}
\date {\today}

\begin{abstract}

In the limit of perfect nesting, the physics of iron-pnictides is governed by
the density wave formation at the zone-edge vector ${\mathbf M}$. At high energies,
various spin- (SDW), charge- (CDW), orbital/pocket-
(PDW) density waves, and their linear combinations, all appear equally likely,
unified within the unitary order parameter of $U(4)\times U(4)$ symmetry.
Nesting imperfections and low-energy interactions reduce
this symmetry to that of real materials. Nevertheless,
the generic ground state preserves a
distinct signature of its highly symmetric origins: a
SDW along one axis of the iron lattice is predicted to {\em coexist} with
a perpendicular PDW, accompanied by weak charge currents.
This ``hidden" order induces the structural
transition in our theory, naturally insures
$T_s \geq T_N$, and leads to orbital ferromagnetism and
other observable consequences.

\end{abstract}
\maketitle


The discovery of high-temperature superconductivity (HTS)
in iron-pnictides \cite{LaOFeAs,BaFe2As2} has sparked
an intense activity \cite{Greene}.
Like the cuprates, the
pnictides are layered systems and exhibit
anti-ferromagnetism (AF) at zero doping ($x=0$), followed
by HTS beyond some finite $x$ \cite{Greene,Mazin}.
Magnetic order in parent compounds consists of
an AF spin chain along the wave vector $(\pi, 0)$
or $(0,\pi)$ in the {\em unfolded} Brilliouin zone (UBZ) and an
FM spin chain along the perpendicular direction \cite{CruzNature2008}.
The dynamical origin of this AF
state is hotly debated: Within the itinerant electron model,
the magnetic transition is ascribed to the
SDW instability, enhanced by the near-nesting
among electron and hole pockets of the Fermi surface (FS)
\cite{Chubukov,VladBand,Wang,AFMIron2008}.
To insure ``striped" spin order, only one electron pocket
is involved in SDW, and the spin-wave anisotropy arises from
the electron pockets' finite ellipticity \cite{EreminMagDeg2010,Knolle}.
In contrast, within the localized Heisenberg-type model
\cite{Si,Bernevig} various
frustrated couplings $J_{1a}$, $J_{1b}$, $J_2$ between neighboring spins
conspire to produce the observed magnetic order
and the magnon anisotropy \cite{JZhaoNat0904,CBroPRL09}.

In addition, the tetrahedral-to-orthorhombic
structural transformation is observed, accompanied by the AF
transition \cite{ST1111,ST122}. The AF ordered  moment is linearly
proportional to orthorhombicity upon change in $x$, and both
transitions disappear for $x> x_c$ \cite{STMTLinear}.
Magnetoelastic coupling was suggested as
being responsible for the close relation
between two transitions \cite{STMTCoupling}. In this approach, the structural
transition is driven by magnetic interactions \cite{Gorkov}. However,
in the 1111 compounds, the structural transition
temperature $T_s$ is consistently above the
AF one, $T_N$, at any $x$ \cite{CruzNature2008}.
Furthermore, the in-plane resistivity anisotropy develops well
above $T_N$ in presence of uniaxial pressure, and
hints at the appearance of a new form of order near $T_s$\cite{ResAnisotropy}.
One possible explanation for $T_s >T_N$ is that magnetic fluctuations are much
stronger than those associated with structural order.

In this Letter, we advance 
another physical picture to account for this evident close relation between the
structural and magnetic transitions: {\em
the two are just different faucets of one
and the same type of ordering of much higher, $U(4)\times U(4)$ symmetry}.
This high symmetry characterizes the dynamics
of pnictides within the {\em high-energy}
regime, extending from the energies of order of the effective bandwidth $D$
down to those set by $T_s\sim T_N$. This regime is governed by ``perfect" nesting
and the tendency toward formation of
a valley-density wave (VDW) at the nesting vector ${\bf Q}$,
with all of its different reincarnations -- various spin-, charge-,
and orbital/pocket-density waves, SDW, CDW, PDW, respectively, as well as
their mutually orthogonal linear combinations -- unified within
a unitary $U(4)\times U(4)$ order parameter \cite{VladDW}.
At yet {\em lower energies}, however,
as the $U(4)\times U(4)$ symmetry-breaking
interactions and the deviations from perfect nesting come
into play, the symmetry is reduced down to that
of real materials. Nevertheless, provided there is a
significant segregation of scales in the effective Hamiltonian of iron-pnictides
between the high-energy $U(4)\times U(4)$-symmetric and the low-energy
symmetry-breaking terms, the ground-state and its excitations bear a distinct
signature of their highly-symmetric origin.

\begin{figure}[htb]
\includegraphics[scale=0.30]{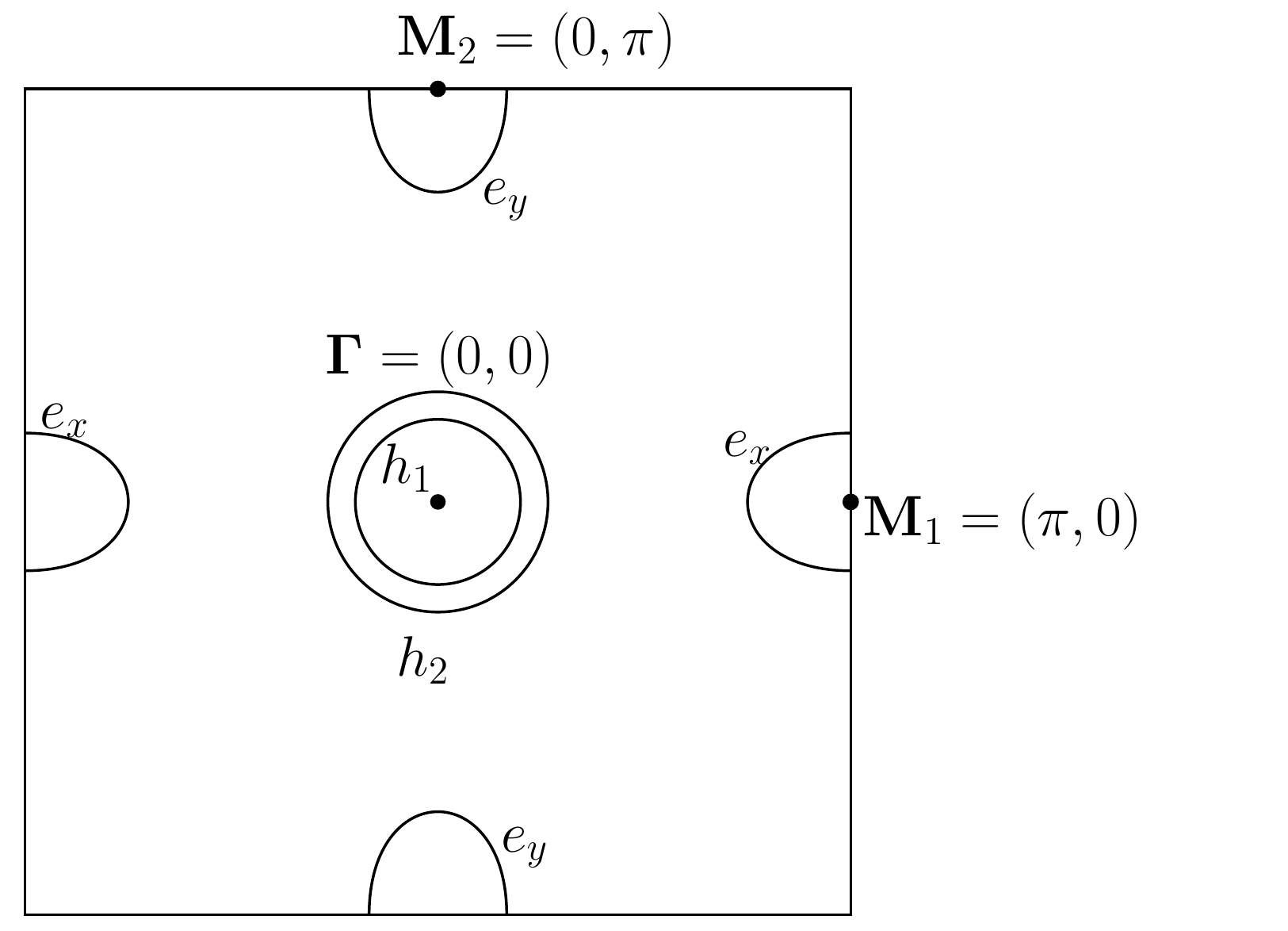}
\caption{Fermi pockets in the UBZ of iron-pnictides.
Two hole pockets $h_1$ and $h_2$ are centered at the $\Gamma = (0 , 0)$ point.
The electron pockets $e_x$ and $e_y$ are centered at the
nesting vectors $\mathbf M_1 = (\pi, 0)$
and $\mathbf M_2 = (0, \pi)$, respectively. The $h_1$, $e_x$, and $e_y$ pockets are
assumed to be perfectly nested to the {\em leading order}, while $h_2$ is
larger than these three; this difference, however, is small
compared to the overall bandwidth $D$, as is the finite but small
ellipticity of $e_x$ and $e_y$ pockets \cite{VladBand,EreminMagDeg2010}.}
\label{Fermi_Pocket_UnBZ}
\end{figure}

Our picture is based on the itinerant model, and relies
on the hierarchy of energy scales that separate the ``flavor"-conserving from
the ``flavor"-changing interactions of quasiparticles on the
FS, composed of two hole ($h_1$, $h_2$) and
two electron ($e_1$, $e_2$) pockets (or valleys) (Fig. 1).
This hierarchy is further assisted by the
differences in area and shape of different pockets
being much smaller than their common overall features;
hence the $U(4)_e \times U(4)_h$ symmetry. Such
hierarchy, quantified in \cite{VladDW}, does not
reflect a deep underlying principle; rather, it is an
accident of the particular semimetallic character of pnictides
and a screened Coulomb repulsion
\cite{footnoteemergence}. But be that as it
may, the hierarchy is well obeyed in all parent compounds and
we use it as an organizing framework to derive the following results:
{\em i}) The ground state of parent pnictides is the {\em combination} of
a SDW along the wave vector $(\pi, 0)$
or $(0,\pi)$ in the UBZ {\em and} a {\em spin-singlet} density wave (DW) along
the perpendicular direction;
{\em ii}) The spin-singlet DW is predominantly
a PDW, with a tiny admixture of a CDW, and is {\em imaginary}, {\em i.e.}
it represents a modulated pattern of weak {\em currents} on
inter-iron bonds. This PDW is difficult to detect
and is dubbed the ``hidden" order;
{\em iii}) The imaginary PDW at ${\bf Q}=(\pi, 0)$ (or $(0,\pi)$)
induces {\em real} CDW at $2{\bf Q} = (0,0)$,
{\em different} from the CDW similarly generated by the SDW.
The resulting broken orbital symmetry between $e_x$ and $e_y$ pockets
(Fig.~\ref{Fermi_Pocket_UnBZ}) drives the observed
tetragonal-to-orthorhombic transition; and
{\em iv}) The predicted electronic structure of the ground state
has numerous observable consequences, some of which we explore.
Our results are generic for the 1111 and 122 materials,
and -- with details changing from one compound to
another -- the overall physical picture should
be universally applicable.

First, we set up the problem:
the band structure can be described by the
five $3d$Fe and three $p$Pn orbitals tight-binding model \cite{VladBand},
resulting in the FS of Fig.~\ref{Fermi_Pocket_UnBZ}.
Our point of departure is the Hamiltonian $H= H_0 + H_{W}$:
\begin{eqnarray}
  H_0 & = & \sum_{\bk, \sigma, \alpha} \epsilon_{\bk}^{\alpha} h_{\bk \sigma}^{(\alpha)\dag} h_{\bk \sigma}^{(\alpha)} + \sum_{\bk, \sigma, \beta} \epsilon_{\bk}^{\beta} e_{\bk \sigma}^{(\beta)\dag} e_{\bk \sigma}^{(\beta)} \nonumber \\
  H_W & = & W \sum_{\bq} \hat n_{\bq}^e \hat n_{-\bq}^h~,
\label{symmetricH}
\end{eqnarray}
where $\sigma, \alpha$, and $\beta$ are the spin,
hole ($h$) and ($e$) pocket indices, respectively (Fig.~\ref{Fermi_Pocket_UnBZ};
$\beta = x, y$ for $e$ bands,
$\alpha = 1, 2$ for $h$ bands) and
$\hat n_{\bq}^e$ and $\hat n_{\bq}^h$ are the density operators
within the $e$ and $h$ pockets \cite{VladDW}.

$H$ (\ref{symmetricH}) describes
the {\em high-energy} physics of pnictides. It contains only the
{\em density-density}, flavor-conserving
interactions between different pockets,
$W\lesssim D$ \cite{footnoteU}. In contrast,
the flavor-changing interactions {\em and}
the variations among $W$s in different pockets are
all $\ll D$, as long as the Hund coupling $J_H\ll U_d$, the Hubbard
repulsion on $d$-orbitals \cite{VladBand,VladDW}.
Furthermore, we also initially assume perfect nesting, {\em i.e.},
$\epsilon_{\bk}^1 = \epsilon_{\bk}^2 = - \epsilon_{\bk + \mathbf M_1}^x = - \epsilon_{\bk + \mathbf M_2}^y = \epsilon_{\bk}$,
since the differences among $h$ and $e$ bands are also $\ll D$.

$H$ (\ref{symmetricH}) has a large $U(4)_e \times U(4)_h$ symmetry,
made manifest by introducing annihilation
operators $c_{\mu}$ and $d_{\nu}$ to represent $h$ and $e$
pockets, respectively, with $\mu, \nu = 1,\dots ,4$ labeling both
spin and band indices:
\begin{align*}
  \mu, \nu  & =  \left\{  \begin{aligned}
 1 \quad   h_{1 \uparrow} \  or \  e_{x\uparrow}; \qquad
 2 \quad   h_{1 \downarrow} \  or \  e_{x\downarrow} \\
 3 \quad   h_{2 \uparrow} \  or \  e_{y\uparrow}; \qquad
 4 \quad   h_{2 \downarrow} \  or \  e_{y\downarrow} \end{aligned} \right. ~. \nonumber \\
 H_0 & =  \sum_{\bk, \mu} \epsilon_{\bk} \left( c_{\mu}^{\dag} c_{\mu} - d_{\mu}^{\dag} d_{\mu} \right) \\
 H_W & =  W \sum_{\bq \bk \mu \nu} c_{\mu \bk + \bq}^{\dag} c_{\mu \bk} d^{\dag}_{\nu \bk'} d_{\nu \bk' + \bq}~.
\end{align*}
The interaction $H_W$ drives a VDW formation at the
nesting vectors $\mathbf M_1$ and $\mathbf M_2$ (Fig. 1). The order
parameter is a $4\times 4$ matrix $\Delta_{\mu\nu}$
whose 16 complex elements describe various SDWs, CDWs, PDWs, and their
linear combinations that gap the FS below some
temperature $T_V$:
\begin{align}
 &\exp \left( - W \sum c_{\mu}^{\dag} c_{\mu} d^{\dag}_{\nu} d_{\nu} \right)
\leftrightarrow \nonumber \\
\int {\cal D} & \Delta\exp \left\{- \sum_{\mu \nu} \left[ \frac1W \vert \Delta_{\mu\nu} \vert^2 -  \Delta_{\mu\nu}^* c^{\dag}_{\mu} d_{\nu} + h.c. \right] \right\} ~.   \nonumber
\end{align}
Integrating out the fermions yields
an effective action ${\cal S}_\Delta$ for bosonic fields
$\Delta_{\mu\nu}$. ${\cal S}_\Delta [\Delta_{\mu\nu}]$ has the
$U(4)_e \times U(4)_h$ symmetry, spontaneously broken at $T_V$. Near
$T_V$, a Ginzburg-Landau (GL) expansion in
$\Delta_{\mu\nu}$ gives \cite{footnoteGL}:
\begin{eqnarray}
T & {\cal S}_\Delta & \to F = \alpha {\mathcal Tr}(\Delta^{\dag} \Delta) + {\tst \half}\beta {\mathcal Tr}(\Delta^{\dag} \Delta \Delta^{\dag} \Delta) \label{F_GL_W}, \\
  \alpha & = & \frac1{W} - \frac{T}N \sum_{k, n} \frac1{\omega_n^2 + \epsilon_k^2} \approx \frac1{W} - N(0) \ln \left( \frac{D}T \right) ~, \nonumber \\
  \beta & = & \frac{T}{2 N} \sum_{k, n} \left( \frac1{\omega_n^2 + \epsilon_k^2} \right)^2 = \frac{7 }{16 \pi^2} \frac{N(0)}{T^2} \zeta(3)~,  \label{Beta_Coef}
\end{eqnarray}
where $N(0)$ is the density of states of a Fermi pocket and $\{\omega_n\}$
are Matsubara frequencies. For $T<T_V$, $\alpha <0$ and
$F$ has a nontrivial minimum for
$\Delta^{\dag} \Delta = \Delta_0^2 = -\alpha /\beta$.
The solution is $\Delta =\Delta_0 {\mathcal U}$, where ${\mathcal U}$ is a $4\times 4$
unitary matrix. At this stage, the four complex 4-vectors comprising ${\mathcal U}$
describe a plethora of SDWs, CDWs, PDWs, etc., and all their
mutually orthogonal linear combinations.
\begin{figure}[htb]
\centering
 \includegraphics[scale=0.25]{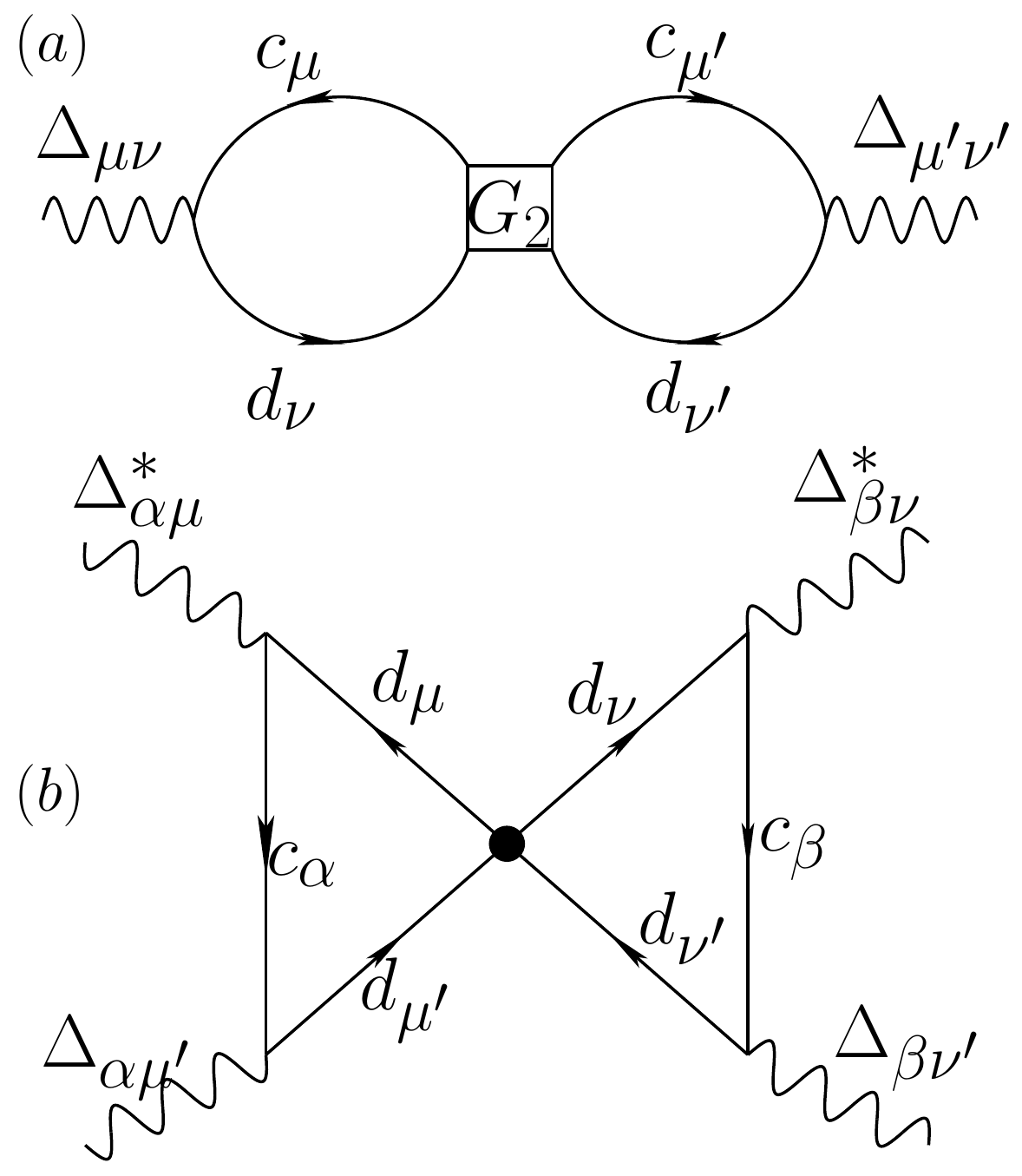} \hspace{0.2cm}
 \includegraphics[scale=0.25]{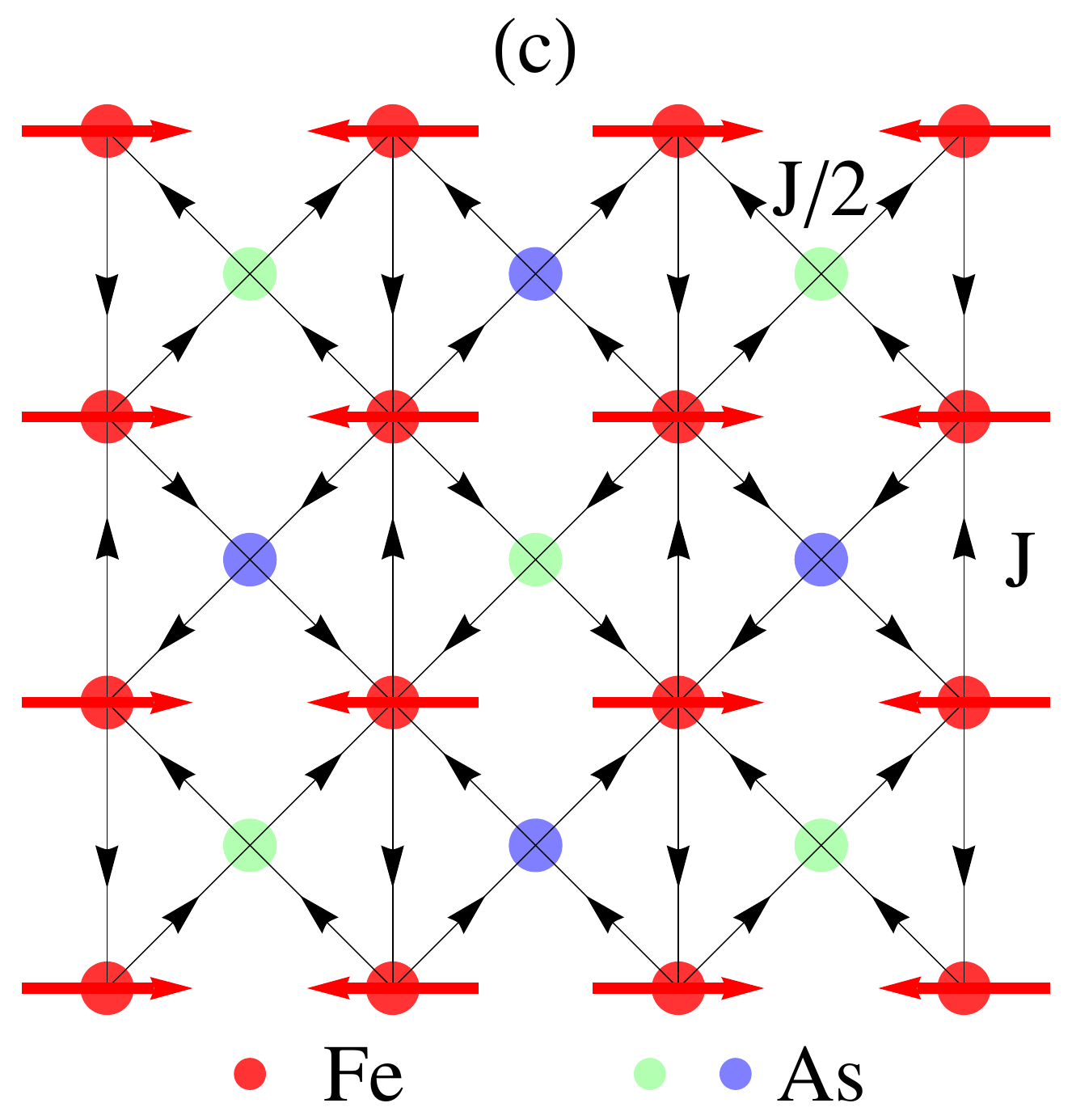}
\caption{(Color) The leading order corrections
to $F$ due to (a) $e-h$ and (b) $e-e$ interactions.
(c) The ground state of parent iron-pnictides. The red and black arrows
depict iron spins and the $[\pm{\mathcal J},\pm\half{\mathcal J}]$ current pattern,
respectively. The ground state {\em combines two}
orders: SDW along $(\pi, 0)$ and the modulated current DW
at the wavevector $(0, \pi)$, {\em i.e.}, the ``hidden" order.}
\label{Correction_GS}
\end{figure}

Now, we are ready to confront the {\em real} iron-pnictides. We
turn on {\em all low-energy} ($\ll D\sim W$) features ignored in
(\ref{symmetricH}) -- differences among $W$s, flavor-changing vertices,
nesting imperfections, and the like \cite{footnoteU} -- and proceed to
systematically decode their
effect on the $U(4)_e \times U(4)_h$
symmetric theory \cite{footnoteleadingorder}.
The most important among these is the interband vertex $G_2$,
which generates the $s^{\pm}$ superconductivity as the nesting subsides
\cite{VladBand, Chubukov}:
\be
  \left[ G_2^{eh_1} c_1^{\dag} c_2^{\dag} ( d_2 d_1 + d_4 d_3 ) + G_2^{eh_2} c_3^{\dag} c_4^{\dag} ( d_2 d_1 + d_4 d_3 ) \right] + h.c.~, \nonumber
\label{G2}
\ee
where $G_2^{eh_\alpha}=G_2^{e_xh_\alpha}=G_2^{e_yh_\alpha}$ \cite{footnoteG2}.
The leading order correction $\Delta F^{G_2}$ (Fig.~\ref{Correction_GS}($a$))
to the free energy $F$ (\ref{F_GL_W}) is
\begin{align*}
  &\sim \Pi(0)^2 \left\{G_2^{eh_1} \left( \Delta_{11} \Delta_{22} + \Delta_{13} \Delta_{24} - \Delta_{12} \Delta_{21} - \Delta_{14} \Delta_{23}  \right)  \right. \\
  & + \left. G_2^{eh_2} \left( \Delta_{31} \Delta_{42} + \Delta_{33} \Delta_{44} - \Delta_{32} \Delta_{41} - \Delta_{34} \Delta_{43}  \right)  +  h.c  \right\},
\end{align*}
$\Pi(0) \approx N(0) \ln \left( \frac{D}T \right)$ \cite{footnoteG2}. The
Cauchy inequality mandates $\Delta F^{G_2} \geq - \Pi(0)^2 \Delta_0^2 \left( \vert G_2^{eh_1} \vert + \vert G_2^{eh_2} \vert \right)$. The equality holds when:
{\em a}) for $G_2^{eh_1} >0$, $\Delta_{22} = -\Delta_{11}^*$, $\Delta_{21} = \Delta_{12}^*$, $\Delta_{24} = - \Delta_{13}^*$, $\Delta_{23} = \Delta_{14}^*$. The
VDW involving the $h_1$ pocket is then
the mixture of {\em real} SDW and {\em imaginary spin-singlet} DW;
{\em b}) for $G_2^{eh_1} <0$, the VDW involving $h_1$ pocket is similarly
the mixture of {\em imaginary} SDW and {\em real spin-singlet} DW.
The same holds for the $h_2$ pocket.
%

Consequently, $G_2$ fixes the phases of different DWs.
One expects that {\em both} $G_2^{eh_1}, G_2^{eh_2}>0$,
as the prerequisite for high $T_c$ $s^{+-}$ superconductivity. Hence,
the ground state of parent compounds
must be composed of either  {\em real} SDW(s) or
{\em imaginary spin-singlet} DW(s); the latter
is a general combination of PDW and CDW,
in the nomenclature of \cite{ZhaiAF2009}.
The real DWs are $\propto \cos(\mathbf M \cdot \br)$, with peaks and
troughs on the iron sites (Fig. \ref{Correction_GS}($c$)).
In contrast, the imaginary spin-singlet DW
breaks time-reversal and lattice translation symmetries along $\mathbf M$,
leading to charge/orbital current DW on iron {\em bonds}.

But which one is it, spin-triplet (SDW) or spin-singlet (PDW/CDW) density wave?
We must consider next the flavor-changing p-h analogue of $G_2$, $G_1$:
$\sum_{\alpha,\beta = 1}^2 G_1^{\alpha \beta}  h^{\dag}_{\alpha\sigma} \left( e_{x\sigma} e_{x\sigma'}^{\dag} + e_{y\sigma} e_{y\sigma'}^{\dag}  \right) h_{\beta\sigma'}$.
$G_1$ ($<G_2$ \cite{VladDW}) generates
$\Delta F^{G_1}$ from a diagram similar to Fig.~\ref{Correction_GS}($a$):
\begin{align}
 \Delta F^{G_1} = & \Pi(0)^2 \left\{ G_1^{11} \left( \vert \Delta_{11} + \Delta_{22} \vert^2 + \vert \Delta_{13} + \Delta_{24} \vert^2 \right)  + \right. \nonumber \\
 & G_1^{22} \left( \vert \Delta_{31} + \Delta_{42} \vert^2 + \vert \Delta_{33} + \Delta_{44} \vert^2 \right) + \nonumber \\
 &  \left[ G_1^{12} (\Delta_{11}^* + \Delta_{22}^*)(\Delta_{31} + \Delta_{42}) + h.c. \right] + \nonumber \\
 & \left. \left[ G_1^{12} (\Delta_{13}^* + \Delta_{24}^*)(\Delta_{33} + \Delta_{44}) + h.c. \right] \right\} \label{F_G1}~.
\end{align}
Here, it is useful to introduce $2\times 2$ $G_1$ matrix
\be
G_1 = \begin{pmatrix}
  G_1^{11} & {\mathcal Re}G_1^{12} \\ {\mathcal Re}G_1^{21} & G_1^{22} \label{Matrix_G1}
\end{pmatrix}~.  \nonumber
\ee
Since the phases of DWs are fixed by $G_2$, only the
real parts of $G_1^{\alpha\beta}$ contribute to $F$.
Hence, $G_1$ is real and symmetric, and has
two real eigenvalues $\lambda_1$, $\lambda_2$, with associated
real eigenvectors $v_1$ and $v_2$. From \eqref{F_G1},
$\Delta F^{G_1}=0$ for SDW and is minimized for
the state composed of:
{\em a}) if $\lambda_1, \lambda_2 >0$,
{\em two} real SDWs;
{\em b}) if $\lambda_1, \lambda_2 <0$, two imaginary spin-singlet DWs,
with $\Delta F^{G_1} = (\lambda_1 + \lambda_2) \Pi(0)^2 \Delta_0^2 < 0$;
and,
{\em c}) if $\lambda_1 < 0$ and $\lambda_2 > 0$, one real SDW
and one imaginary spin-singlet DW,
with $\Delta F^{G_1} = \lambda_1 \Pi(0)^2 \Delta_0^2 < 0$.
Experimentally, there is only a single SDW at $(\pi, 0)$.
This implies option {\em c}): with majority of $G_1$s
rather small \cite{VladDW}, this is to be expected, once
we include the (weak) electron-phonon
coupling \cite{Greene,Mazin} 
and large polarizability
of pnictide $p$ orbitals \cite{VladBand}. 
In this case, the leading order contribution
of $\Delta F^{G_2}$ and $\Delta F^{G_1}$ to $F$ is
\be
F \approx \alpha (\Delta_{\rm SDW}^2 + \Delta_{\rm SSDW}^2) + \lambda_1 \Pi(0)^2 \Delta_{\rm SSDW}^2~. \label{Approx_F} \nonumber
\ee
$\Delta_{\rm SDW}$ and $i \Delta_{\rm SSDW}$ describe
the SDW and the imaginary spin-singlet DW, respectively, while
\begin{align*}
& \alpha(T_{\rm SDW}) = 0 \qquad \alpha(T_{\rm SSDW}) + \lambda_1 \Pi(0)^2 = 0 ~, \\
& \frac{T_{\rm SSDW} - T_{\rm SDW}}{T_{\rm SDW}} \approx -\frac{\lambda_1/W}{N(0)W} > 0 ~,
\end{align*}
set the corresponding transition temperatures.
As long as $\vert \lambda_1 \vert \ll W$ \cite{VladDW},
$T_{\rm SSDW} \gtrapprox T_{\rm SDW}$ and $\Delta_{\rm SSDW} \gtrapprox \Delta_{\rm SDW}$.

Consider now $v_1 = (a, b)$, the (real) eigenvector associated with
$\lambda_1 <0$. $\Delta F^{G_1}$ is minimized by
\begin{align}
\Delta(\theta) & = \Delta_0 \begin{pmatrix}  i a {\mathbb 1} &  - b\sigma_n
\\ i b {\mathbb 1} &  a \sigma_n \end{pmatrix} \times
\begin{pmatrix}
  \cos\theta {\mathbb 1} & - \sin\theta {\mathbb 1} \\ \sin\theta {\mathbb 1} & \cos\theta {\mathbb 1}
\end{pmatrix}~,
\label{Delta_IsoSpin}
\end{align}
where $\sigma_n = \vec \sigma \cdot \hat n$,
$\hat n$ is an arbitrary unit vector reflecting the
$SU(2)$ spin symmetry of our theory, and $\theta$ is an arbitrary angle,
signaling an additional degeneracy in the
Hamiltonian. The second matrix in \eqref{Delta_IsoSpin}
is a rotation by $\theta$ which mixes $e_x$ and $e_y$ pockets:
\be
e_1 = \cos\theta e_x - \sin\theta e_y~,~e_2 = \cos\theta e_y + \sin\theta e_x~.
\ee
In the state described by \eqref{Delta_IsoSpin}, $e_1$ and $e_2$
couple to $h_n=a h_1 + b h_2$ and $h_p = a h_2 - b h_1$,
respectively, to form two DWs.
Finally, this remaining $\theta$-degeneracy is lifted by
the density-density repulsion between $e_x$ and $e_y$ pockets:
\be
  W_k^e e_{x \sigma}^{\dag} e_{x\sigma} e_{y \sigma'}^{\dag} e_{y\sigma'} \rightarrow  W_k^e \left( d_1^{\dag} d_1 + d^{\dag}_2 d_2 \right) \left( d_3^{\dag} d_3 + d^{\dag}_4 d_4 \right)~, \nonumber
\ee
with $W_k^e > 0$. The leading order contribution to $F$,
$\Delta F^{W_k}$, follows from
Fig.~\ref{Correction_GS}($b$),
and contains two fermion loops, each with three legs.
Were the nesting perfect,
the loop integral would be independent of leg indices, and, upon
summation over hole indices, the contribution of each loop would be
$\propto\Delta^{\dagger} \Delta$, but still {\em independent} of $\theta$.

In {\em real} pnictides, however, the outer pocket $h_2$
deviates significantly from $h_1$ and perfect
nesting (Fig.~\ref{Fermi_Pocket_UnBZ})\cite{Greene,VladBand,FP_Exp_Iron}.
To account for this, we set $\epsilon^{h_2}_{\bk} = \epsilon^{h_1}_{\bk} + \eta$,
$\eta \ll W\lesssim D$. At the leading order in $\eta$,
the $\theta$-dependent term of each fermion
loop in Fig.~\ref{Correction_GS}($b$) is now finite and contributes
\begin{align}
\frac1N \sum_{\omega, \bk} & \left( \frac1{i \omega + \epsilon} \right)^2 \frac{\eta}{ \left( i \omega - \epsilon \right)^2 } = 2 \beta \eta  \to  \nonumber \\ 
\to \Delta  F^{W_k} &  \sim 2 W_k^e (2 \beta \eta)^2 \Delta_0^2 [(a\cos\theta)^2 + (b \sin\theta)^2] \times \nonumber \\
\quad [(a\sin\theta)^2  + & (b  \cos\theta)^2]
\propto  (ab)^2 + (b^2 - a^2)^2 \cos^2\theta \sin^2\theta~.\nonumber
\end{align}
Since generally $ \vert a \vert \not = \vert b \vert$, $\Delta F^{W_k}$ is minimized
for $\theta = 0$ or $\pi/2$. {\em Thus, the preferred ground state
combines a real SDW in one direction and an
imaginary spin-singlet DW along the perpendicular direction}. The nature
of this imaginary spin-singlet DW depends on the form of
$v_1$ \cite{footnoteW}. If $a \approx -b$, the spin-singlet DW is
predominantly a PDW, translating into a purely {\em orbital} current
pattern. However, unless $a = -b$, there is also an accompanying
{\em charge} current DW, depicted in Fig.~\ref{Correction_GS}($c$). This
current DW can be weak for generic $a\sim -b$ but should be observable
and is the main prediction of this Letter.
Since the charge current DW interacts with the underlying
lattice more strongly than the pure PDW, it favors
an additional modulated structural
pattern along $(0, \pi)$, on top of the one tied to
the SDW along $(\pi, 0)$. The apparent absence of such pattern in
pnictides suggests that indeed $a\approx -b$ and the PDW dominates the
imaginary spin-singlet DW.

With two DWs present at ${\mathbf Q} = \bf M_1$ and $\bf M_2$, a real CDW at $2{\mathbf Q} = (0,0)$ is {\em induced} as a next harmonic \cite{balatsky}. First, this is illustrated within a two-band model, with one $h$ and one $e$ pocket. In the mean-field approximation:
\begin{align}
& H_{MF} = \delta\Sigma (h^{\dag}_{\bk \sigma} h_{\bk \sigma} -e^{\dag}_{\bk \sigma} e_{\bk \sigma});H_U  = U (\hat n_h^2 + \hat n_e^2), \nonumber \\
& F \leq F_{MF} + \langle H_U - H_{MF} \rangle_{MF}   ~.\label{F_CDW}
\end{align}
$U$ is the intrapocket repulsion and $\delta\Sigma$ is the relative shift of $h$ and $e$ self-energies. Here we assume the $e$ pocket dispersion is $\epsilon_{\bk} = k^2/2m - \epsilon_0$. For $\delta\Sigma \ll \Delta_0$,
\begin{align*}
  & F_{MF}  =  \alpha(\delta\Sigma) {\mathcal Tr}(\Delta^{\dag} \Delta) + O(\Delta^4), \\
  & \alpha(\delta\Sigma) =  \frac1W - \frac1\beta \sum_n \int^{D}_{-\epsilon_0 - \delta\Sigma} \mathrm{d}\epsilon \frac{N(0)}{\epsilon^2 + \omega_n^2} \approx \alpha(0) - \delta\Sigma \frac{N(0)}{2 \epsilon_0},\nonumber \\
  & \langle \delta n_e \rangle = - \langle \delta n_h \rangle \approx N(0) \delta\Sigma,\\
  & \langle H_U - H_{MF} \rangle_{MF} \approx 2 N(0) \left( \delta\Sigma \right)^2 ( 1 + N(0)U )~.
\end{align*}
The r.h.s. of \eqref{F_CDW} is minimized when $\delta \Sigma = \Delta_0^2/(8 \epsilon_0 (1 + N(0) U))$, and hence,
$\langle \delta n_e \rangle = - \langle \delta n_h \rangle =  \frac{N(0) \Delta_0^2 }{8 \epsilon_0 (1 + N(0) U)}$.

In a realistic four band model, with the induced CDWs at $2{\mathbf Q}$, a lengthy but
straightforward algebra yields \cite{footnoteleadingorder}
\begin{eqnarray}
  \langle \delta n_{e_x} \rangle & = & - \langle \delta n_{h_p} \rangle  = \frac{N(0)}{8 \epsilon_0(1 + N(0)U)} \Delta_{\rm SDW}^2~, \nonumber \\
  \langle \delta n_{e_y} \rangle & = & - \langle \delta n_{h_n} \rangle  = \frac{N(0)}{8 \epsilon_0(1 + N(0)U)} \Delta_{\rm PDW}^2~.
\label{occupancies}
\end{eqnarray}
As shown earlier, $\Delta_{\rm SDW} < \Delta_{\rm PDW}$, and thus $\langle e_y^{\dag} e_y \rangle > \langle e_x^{\dag} e_x \rangle$ 
implying unequal occupancy and splitting
of $d_{xz}$ and $d_{yz}$ orbitals. 
Consequently, the induced real CDW at $2{\mathbf Q} = (0,0)$ 
is an orbital ferromagnet which
breaks the $C_4$ symmetry while preserving the lattice translation symmetry, and can be naturally identified as the source of the observed tetragonal-to-orthorhombic distortion. Since the CDW arises simultaneously with the modulated DWs, $T_s = T_{\rm PDW} \geq T_N= T_{\rm SDW}$. For $T\ll T_{\rm SDW}$ and $0<x\ll x_c$, Eqs.~(\ref{occupancies}) also result in orthorhombicity $\propto \Delta_{\rm SDW}(x)$, in agreement with \cite{STMTLinear}.
Additional support for this picture of structural deformation
comes from the universal scaling of magnetization \cite{birgeneau}.

In summary, we have shown that the high-energy $U(4)\times U(4)$ symmetry in iron-pnictides naturally leads to the prediction of a ``hidden" orbital current DW order in parent compounds 
near the $\lambda_1 = 0$ quantum critical point
and have explored some of the observable consequences.

\begin{acknowledgments}

We thank V. Cvetkovic for discussions and for
sharing his insights with us. This work was supported in part by the Johns Hopkins-Princeton Institute for Quantum Matter, under Award No.\ DE-FG02-08ER46544 by the U.S.\ Department of Energy, Office of Basic Energy Sciences, Division of Materials Sciences and Engineering.

\end{acknowledgments}






\begin{thebibliography}{1}
\bibitem{footnoteU} The intraband repulsion is already implicitly included in (\ref{symmetricH}) through renormalized bandwidths and other harmless Fermi-liquid renormalizations.
\bibitem{footnoteGL} Note that the GL expansion is by no means necessary and is used only to make our presentation manageable.
\bibitem{footnoteleadingorder} For clarity, this is done to the leading order. The basic results, however, are robust and remain valid to all orders, as long as the energy hierarchy remains in place, as is evident from the text.
\end{thebibliography}
\end {document}